\documentclass[a4paper]{article}

\usepackage{INTERSPEECH2022}

\usepackage{amsmath,graphicx}
\usepackage[cmintegrals]{newtxmath}
\usepackage{bm}
\usepackage{algorithmicx,algorithm}
\usepackage{booktabs}
\usepackage{multirow}
\usepackage{graphicx}
\usepackage{stfloats}
\usepackage{subfig}
\usepackage{xcolor}
\usepackage{hyperref}
\usepackage{xspace}
\usepackage{enumitem}
\usepackage{bbding}
\usepackage{verbatim}

\hypersetup{hidelinks}

\def\equationautorefname#1#2\null{
	Eq. (#2\null)
}

\newcommand{\method}{DecoupleFL\xspace}

\title{Decoupled Federated Learning for ASR with Non-IID Data}
\name{Han Zhu$^{1,2}$, Jindong Wang$^{3}$, Gaofeng Cheng$^{\dagger1}$\thanks{$\dagger$ Corresponding author.}\thanks{This work is partially supported by the Youth Innovation Promotion Association, Chinese Academy of Sciences and the Frontier Exploration Project Independently Deployed by Institute of Acoustics, Chinese Academy of Sciences under Grant QYTS202011.}, Pengyuan Zhang$^{1,2}$, Yonghong Yan$^{1,2}$}
\address{$^1$Key Laboratory of Speech Acoustics and Content Understanding, Institute of Acoustics CAS, China \\
$^2$ University of Chinese Academy of Sciences, China  \qquad  $^3$ Microsoft Research Asia, China}
\email{\{zhuhan,chenggaofeng,zhangpengyuan,yanyonghong\}@hccl.ioa.ac.cn,jindong.wang@microsoft.com}

\begin{document}
%
\maketitle
\begin{abstract}
Automatic speech recognition (ASR) with federated learning (FL) makes it possible to leverage data from multiple clients without compromising privacy. The quality of FL-based ASR could be measured by recognition performance, communication and computation costs. When data among different clients are not independently and identically distributed (non-IID), the performance could degrade significantly. In this work, we tackle the non-IID issue in FL-based ASR with \emph{personalized FL}, which learns personalized models for each client. Concretely, we propose two types of personalized FL approaches for ASR. Firstly, we adapt the \emph{personalization layer based FL} for ASR, which keeps some layers locally to learn personalization models. Secondly, to reduce the communication and computation costs, we propose \emph{decoupled federated learning (\method)}. On one hand, \method moves the computation burden to the server, thus decreasing the computation on clients. On the other hand, \method communicates secure high-level features instead of model parameters, thus reducing communication cost when models are large. Experiments demonstrate two proposed personalized FL-based ASR approaches could reduce WER by 2.3\% - 3.4\% compared with FedAvg. Among them, \method has only 11.4\% communication and 75\% computation cost compared with FedAvg, which is also significantly less than the personalization layer based FL.
\end{abstract}

\section{Introduction}
\label{sec:intro}

\begin{figure}[!t]
\centering
	\subfloat[Centralized]
	{ \label{fig:compare-a}
		\includegraphics[width=0.28\columnwidth]{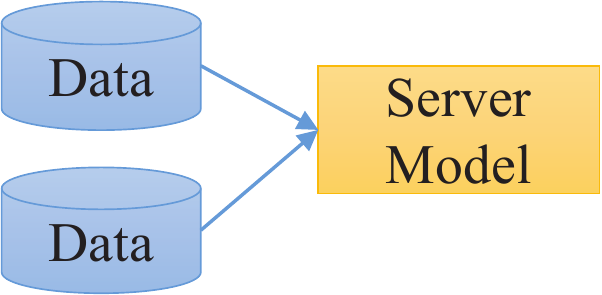}
	}		
	\hfill
	\subfloat[FedAvg]  
	{ \label{fig:compare-b}
		\includegraphics[width=0.45\columnwidth]{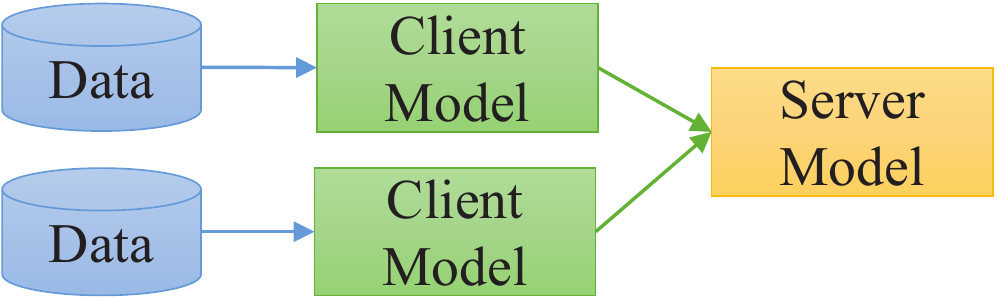}
	}
	\newline
	\subfloat[\method]  
	{ \label{fig:compare-c}
		\includegraphics[width=0.6\columnwidth]{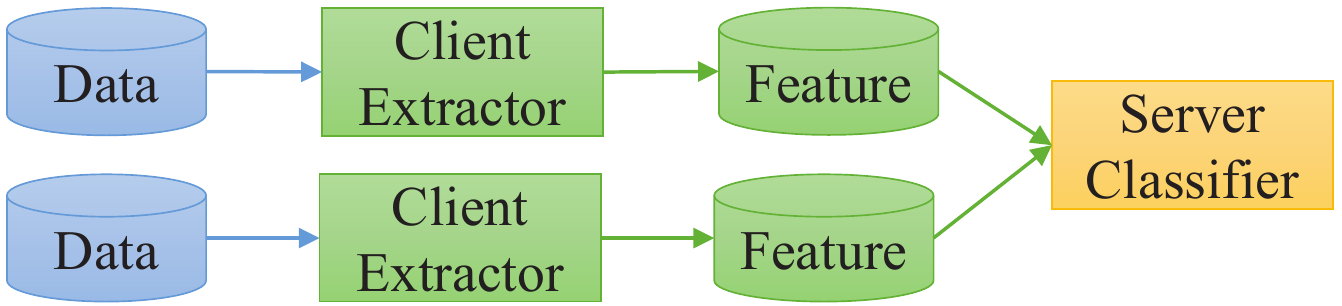}
	}
	\label{fig:compare}
	\caption{Illustration of different training methods.} 
\end{figure}

ASR relies on massive training data for decent performance and conventionally uses centralized training as in \autoref{fig:compare-a}, where raw data of all clients are aggregated to the server.
However, due to concerns and regulations \cite{voigt2017eu} of data privacy, the client's data is in the form of isolated islands and is not allowed to be shared. Therefore, federated learning (FL) \cite{mcmahan2017communication,chen2020fedhealth} is proposed to collaboratively train the model for many clients without compromising privacy under the coordination of a server.
Existing literature on FL-based ASR \cite{dimitriadis2020federated,tan2020novo,cui2021federated,guliani2021training,gao2021end,nandury2021cross,yang2021partial,guliani2021enabling} mostly follows the paradigm of FedAvg \cite{mcmahan2017communication}, where model parameters are exchanged instead of raw data.
As shown in \autoref{fig:compare-b}, client models are trained locally for some epochs (local epoch) and then aggregated globally in the server. 
The local training and global aggregation are performed multiple times (global epoch).

FedAvg trains a global model for all clients and ignores the personalization of each client. Therefore, its performance could degrade when data among clients are non-IID~\cite{zhao2018federated}. Personalized FL \cite{mansour2020three} can be used to alleviate this issue by learning personalized models for individual clients. Local fine-tuning based approaches study how to obtain a better initial global model \cite{fallah2020personalized, jiang2019improving} or perform elaborate local optimization \cite{yu2020salvaging}. Model mixture approaches \cite{deng2020adaptive,mdhaffar2021study} propose to interpolate the local and the global models. Personalization layer based approaches
\cite{li2020fedbn,arivazhagan2019federated,liang2020think,collins2021exploiting} keep some layers locally for personalized models while aggregating the rest with FedAvg. In this work, we adapt this idea and design two variants for ASR.

However, existing personalized FL approaches have several limitations: (1) model parameters are communicated between server and clients, leading to high communication costs when models are large; (2) the low-resource client are required to do most computation while the computation on the high-resource server could be neglected; (3) clients data are required to be labeled despite it is troublesome. To tackle these challenges, we propose \emph{decoupled federated learning (\method)} for personalized FL in ASR.
As shown in \autoref{fig:compare-c}, \method decouples the training of the ASR model: the extractor that has contact with raw data is trained on clients (stage 1), and the classifier is trained on the server with the secure features from clients (stage 2). In this way, \method communicates secure features instead of model parameters, thus reducing the communication cost. Then, some training burdens are moved to the server, thus reducing computation on clients. Additionally, \method adopts pseudo-labeling (PL) approaches \cite{higuchi2021momentum,manohar2021kaizen} for unsupervised learning, avoiding the unrealistic labeled data assumption. Moreover, one potential concern is communicating features might lead to privacy leakage. To address such concern, we remove the speaker information from features with speaker-invariant training (SIT) \cite{meng2018speaker,srivastava2019privacy}, which could protect speaker information while does not hamper performance.

Experiments show that compared with FedAvg, all personalized FL approaches (personalization layer based approaches and \method) can reduce WER by 2.3\% - 3.4\%. Among them, \method achieves the lowest communication and computation costs, which are 11.4\% and 75\% of FedAvg. 

\section{Proposed Approach}
\label{sec:proposed}

In this section, we first introduce FL with personalization layers approach and two proposed variants for ASR. Then, we describe the proposed \method approach.

\begin{figure*}[!t]
    \centering
	\subfloat[FL with personalization layers.] 
	{ \label{fig:perfl}
		\includegraphics[width=0.55\columnwidth]{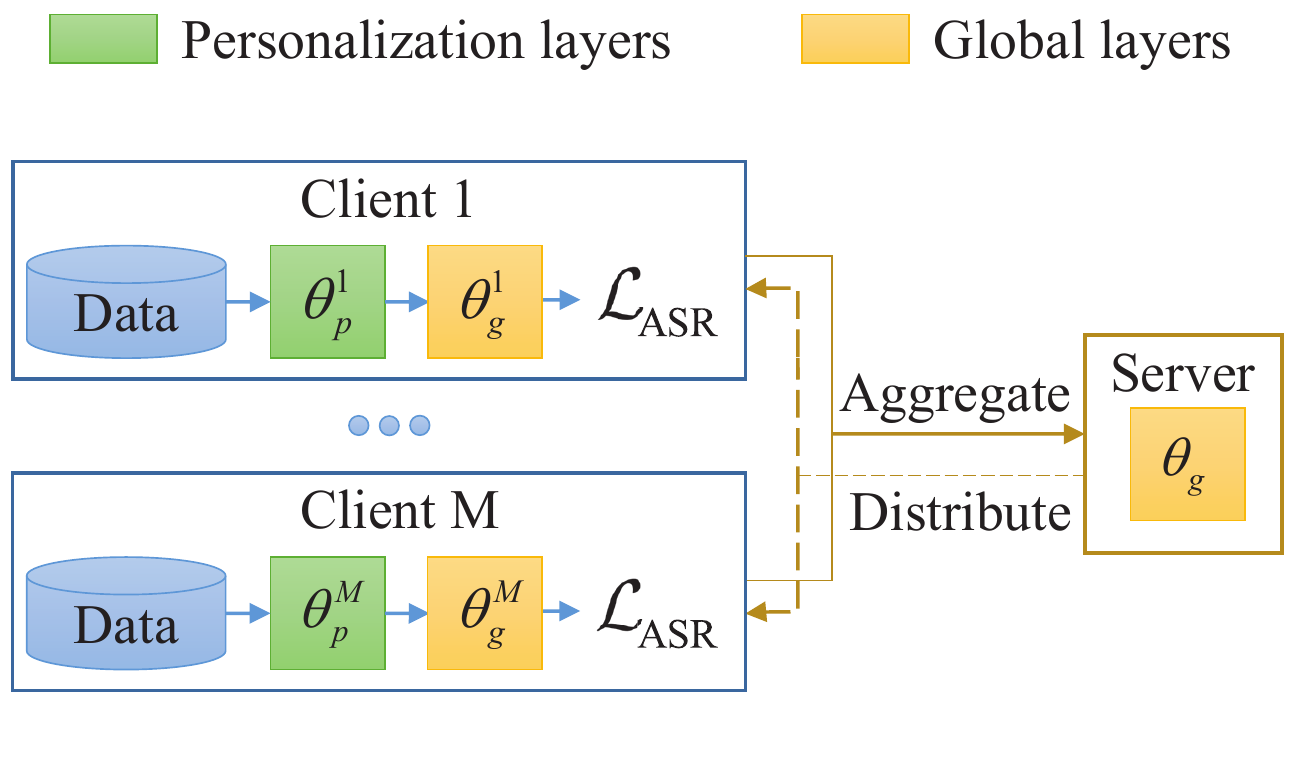}
	}
    \quad
	\subfloat[Decoupled federated learning.] 
	{ \label{fig:DecoupleFL}
		\includegraphics[width=0.7\columnwidth]{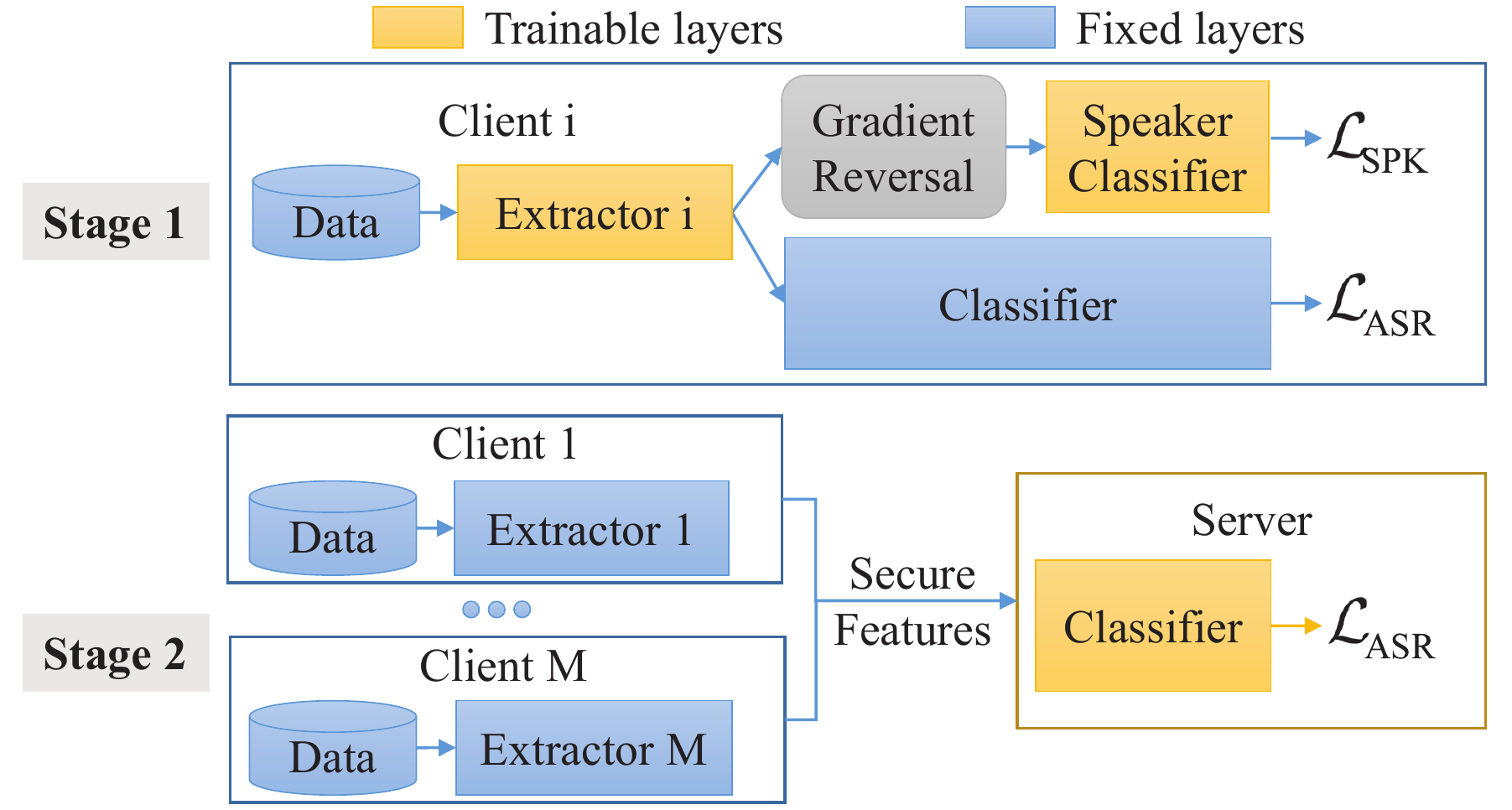}
	}
	\quad
	\subfloat[Continuous PL]  
	{ \label{fig:PL}
		\includegraphics[width=0.4\columnwidth]{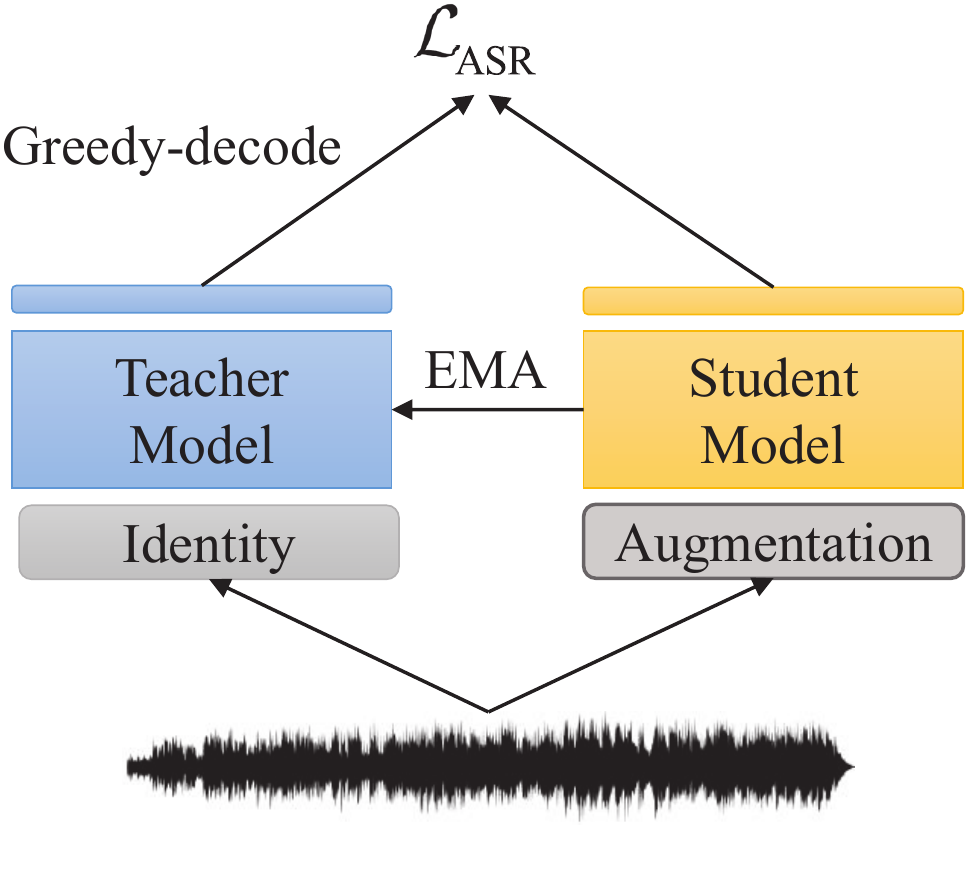}
	}
	\label{fig:decouple}
	\caption{(a) FL with personalized layers. Personalization layers are kept locally while global layers are aggregated in the server. (b) \method. In stage 1, extractors are trained locally with client's data. In stage 2, the classifier is trained globally with secure features from clients. (c) Continuous pseudo-labeling. Pseudo labels are generated from teacher model, which is the EMA of the student model.} 
\end{figure*}

\subsection{FL with Personalization Layers}

A straightforward way for personalized FL is to keep some personalization layers locally and only aggregate other global layers as in \autoref{fig:perfl}. Therefore, the personalization layers can learn the personalized perspective of clients, while global layers learn the common knowledge shared among clients. 

We denote the parameters of personalization and global layers as $\theta_{p}$ and $\theta_{g}$, respectively. Client's index is $k$, training update is $t$, and local update number is $E$. Same with FedAvg, in each client, all layers are trained using stochastic gradient descent (SGD):
\begin{equation}
    \left(\theta^{k}_{p,t+1}, \theta^{k}_{g,t+1}\right) \gets SGD\left(\theta^{k}_{p,t}, \theta^{k}_{g,t}\right).
\end{equation} 

Then, when $mod\left(t, E\right) = 0$, the server aggregates global layers from all $K$ clients and send them back to each client as:
\begin{equation}
    \theta^{k}_{g,t+1} \gets \frac{1}{K} \sum\nolimits^{K}_{k=1} \theta^{k}_{g,t+1}.
\end{equation} 

It is intuitive to design two variants for ASR: FedNorm and FedExtract. FedNorm uses all normalization layers as personalization layers to reduce the distribution shift before each layer; FedExtract uses the bottom few layers (extractor) as personalization layers to reduce the distribution shift before the global classifier. However, they require communicating parameters, leading to high communication costs for large models. Since SGD is performed only on clients while parameter aggregation is performed on the server, the computation burden for low-resource clients is much larger than the high-resource server.

\subsection{Decoupled Federated Learning}
This paper proposes \emph{\method} to address the above challenges in personalized FL for ASR. Specifically, \method decouples the model training into two stages: training of extractor on clients (stage 1) and classifier on the server (stage 2). Thus, it reduces the computation cost on clients and reduces the communication cost when ASR models are large.
The procedure of \method is shown in \autoref{fig:DecoupleFL}.
Note that \method requires a pre-trained ASR model as the initialization.

\subsubsection{Stage 1: Training of Extractor}
\label{sec:stage1}

In stage 1, the classifier $\theta_{c}$ is fixed and the extractors $\theta_{e}$ are optimized locally on client's data. Therefore, similar with FedExtract, extractors are also used as personalization layers to tackle the non-IID distributions among clients. To remove speaker information from the features, we applied speaker-invariant training on the extractor. Specifically, we add a speaker classifier $\theta_{s}$ and train the entire model with the min-max objective:
\begin{equation}
\min _{\theta_{e}} \max _{\theta_{s}} \mathcal{L}_{\text{ASR}}\left(\theta_{e}, \theta_{c}\right)-\lambda \mathcal{L}_{\text{SPK}}\left(\theta_{e}, \theta_{s}\right)
\end{equation}
Where $\mathcal{L}_{\text{ASR}}$ is the ASR loss (CTC) and  $\mathcal{L}_{\text{SPK}}$ is the speaker classification loss (cross-entropy). 

With this objective, the speaker classifier $\theta_{s}$ is optimized to minimize the speaker classification loss, while the extractor $\theta_{e}$ maximizes it. In this way, the extractor is trained to remove speaker information from features so that the speaker classifier cannot accurately classify speakers. Simultaneously, the extractor is optimized to minimize the ASR loss $\mathcal{L}_{\text{ASR}}$, which ensures the extractor could maintain decent ASR performance. This min-max optimization is implemented by inserting a gradient reversal layer between the extractor and speaker classifier \cite{ganin2015unsupervised}.

\subsubsection{Stage 2: Training of Classifier}
\label{sec:stage2}
In stage 2, each client's features extracted with the personalized extractor from stage 1 are aggregated in the server to refine the common classifier. 
Since the personalized extractors are constrained by a fixed common classifier $\theta_{c}$ in stage 1, these features are well aligned. Thus it is reasonable to use these features to refine the common classifier. The optimization objective is:
\begin{equation}
\min _{\theta_{c}}  \mathcal{L}_{\text{ASR}}\left(\theta_{c}\right)
\end{equation}
where the classifier is optimized and extractors are not used.
 
\subsubsection{Unsupervised Training with Continuous PL}
\label{sec:unsupervised}
We adopt continuous PL (shown in \autoref{fig:PL}) \cite{higuchi2021momentum,manohar2021kaizen} to compute the $\mathcal{L}_{\text{ASR}}$ in both stage 1 and stage 2. Note that the continuous PL approach could also be used in other FL approaches like FedNorm and FedExtract.
Given an unlabeled sample $x$, the ASR loss in the $t$-th round of update is computed as:
\begin{equation}
\label{equ:unlabel}
\mathcal{L}_{\text{ASR}}\left(\theta\right) = -\mathbb{E}_{\mathbf{x} \sim p(\mathbf{x})} \log p_{\mathbf{\theta}_{t}}(\hat{\mathbf{y}} \mid a(\mathbf{x})),
\end{equation}
where $\theta_{t}$ is the $t$-th student model, $a(\cdot)$ is the data augmentation function and $\hat{\mathbf{y}}$ denotes the pseudo label which is generated as:
\begin{equation}
\label{equ:pseudolabel}
\hat{\mathbf{y}}=\underset{\mathbf{y}}{\operatorname{argmax}} \log p_{\xi_{t}}(\mathbf{y} \mid \mathbf{x}),
\end{equation}
where $\operatorname{argmax}$ denotes the greedy decoding.
$\xi_{t}$ is the teacher model in the $t$-th update, which is the exponential moving average of the student model $\theta_{t}$:
\begin{equation}
 \xi_{t} = \alpha \xi_{t-1} + (1-\alpha) \theta_{t}  
\end{equation}
where $\alpha$ is the decay factor.

\section{Experiments}
\label{sec:experiments}

\subsection{Experimental Setup}

We evaluate all approaches by adapting a baseline ASR model to an unseen target domain. The baseline ASR model consists of two layers of CNN and 14 layers of transformer. 
We use CNN and the bottom 7 transformer layers as extractors, while other layers are the classifier. The baseline ASR model is centralized trained in the server with CTC criterion on datasets in \autoref{tab:datasets}.

\begin{table}[htbp]
  \centering
  \caption{Structure of labeled datasets for baseline model.}
  \resizebox{.25\textwidth}{!}{
    \begin{tabular}{cc}
    \toprule
    Dataset & Duration (Hours) \\
    \midrule
    AMI & 100 \\
    Fisher & 1,761 \\
    SwitchBoard & 317 \\
    LibriSpeech & 960 \\
    Wall Street Journal (WSJ) & 81 \\
    TED-LIUM v3 & 452 \\
    Internal Dataset & 3,248 \\
    \midrule
    Total & 6,919 \\
    \bottomrule
    \end{tabular}%
    }
  \label{tab:datasets}%
\end{table}%

The target domain dataset is constructed from the unseen Common Voice corpus \cite{ardila2020common}. Specifically, we select three accents from Common Voice: Australia (AU), England (EN), and India (IN), with a training/validation/test ratio of $100h:10h:10h$.
Each accent accounts for 1/3 in each set. We take each accent as a local client to simulate the non-IID scenario.

All methods are trained for 100 epochs using Adam optimizer with the learning rate $10^{-4}$ and decay factor $\alpha = 0.9998$. For evaluation, we average the 10 best checkpoints and apply beam-search decoding with LM, where the 4-gram LM is trained with transcripts in \autoref{tab:datasets}.

\subsection{Unsupervised Training Approach}

\begin{table}[htbp]
  \centering
  \caption{Comparison of PL approaches.}
  \resizebox{.25\textwidth}{!}{
    \begin{tabular}{lcccc}
    \toprule
    Method &  AU & EN & IN & AVG \\
    \midrule
    Baseline & 19.4 & 15.7 & 22.6 & 19.2 \\
    Vanilla PL  & 17.6 & 15.2 & 20.6 & 17.8 \\
    Continuous PL & \textbf{16.9} & \textbf{14.6} & \textbf{19.7} & \textbf{17.1} \\
    \bottomrule
    \end{tabular}%
    }
  \label{tab:compare_pl}%
\end{table}%

We compare continuous PL with vanilla PL \cite{park2020improved,xu2020iterative}, where pseudo labels are generated with a fixed ASR model through beam-search with LM. As shown in \autoref{tab:compare_pl}, continuous PL consistently outperforms vanilla PL in centralized training. Thus, we use continuous PL for unsupervised learning as follows.

\subsection{Main Results}
We compare FedNorm, FedExtract and \method with:
\begin{itemize}[noitemsep]
    \item \emph{Baseline:} the unadapted baseline model.
    \item \emph{Client:} separately adapts baseline model for each client.
    \item \emph{Centralized:} centralized training.
    \item \emph{FedAvg:} the standard FL, where the local epoch is 1.
\end{itemize}

\begin{table}[htbp]
  \centering
  \caption{Comparison of non-federated and federated learning approaches in terms of communication (comm.), computation cost, and word error rate (WER).}
  \resizebox{0.48\textwidth}{!}{
    \begin{tabular}{lrrccccc}
    \toprule
    \multirow{2}[4]{*}{Method} & \multirow{2}[4]{*}{Comm.} & \multicolumn{2}{c}{Computation} & \multicolumn{4}{c}{WER (\%)} \\
\cmidrule{3-8}        &     & client & server & Australia & England & Indian & AVG \\
    \midrule
    Baseline & -   & -   & -   & 19.4 & 15.7 & 22.6 & 19.2 \\
    Client & -   & 100 & -   & 17.4 & 14.9 & 20.3 & 17.5 \\
    Centralized & -   & -   & 100 & 16.9 & 14.6 & 19.7 & 17.1 \\
    \midrule
    FedAvg & 20.70GB & 100 & -   & 17.3 & 14.9 & 20.4 & 17.5 \\
    FedNorm & 20.69GB & 100 & -   & 16.9 & 14.6 & 19.7 & 17.1 \\
    FedExtract & 10.71GB & 100 & -   & 16.8 & 14.5 & 19.8 & 17.0 \\
    DecoupleFL & \textbf{2.35GB} & \textbf{50} & \textbf{0.5 * 50} & \textbf{16.5} & \textbf{14.5} & \textbf{19.7} & \textbf{16.9} \\
    \bottomrule
    \end{tabular}%
    }
  \label{tab:main}%
\end{table}%

As shown in \autoref{tab:main}, centralized outperforms client, indicating more training data helps even when non-IID.
FedAvg only performs similarly with client due to the non-IID issue. FedNorm, FedExtract and \method outperform FedAvg by alleviating the non-IID issue with personalized models. Among them, \method requires much less communication cost (i.e., communicated data size) and decreases computation cost (i.e., training epochs) by 50\% for clients and by 25\% in total, which will be further discussed in \autoref{sec:discussion}.

\begin{figure}[htbp]
\centering
	\subfloat[FedNorm]  
	{  \includegraphics[width=0.3\columnwidth]{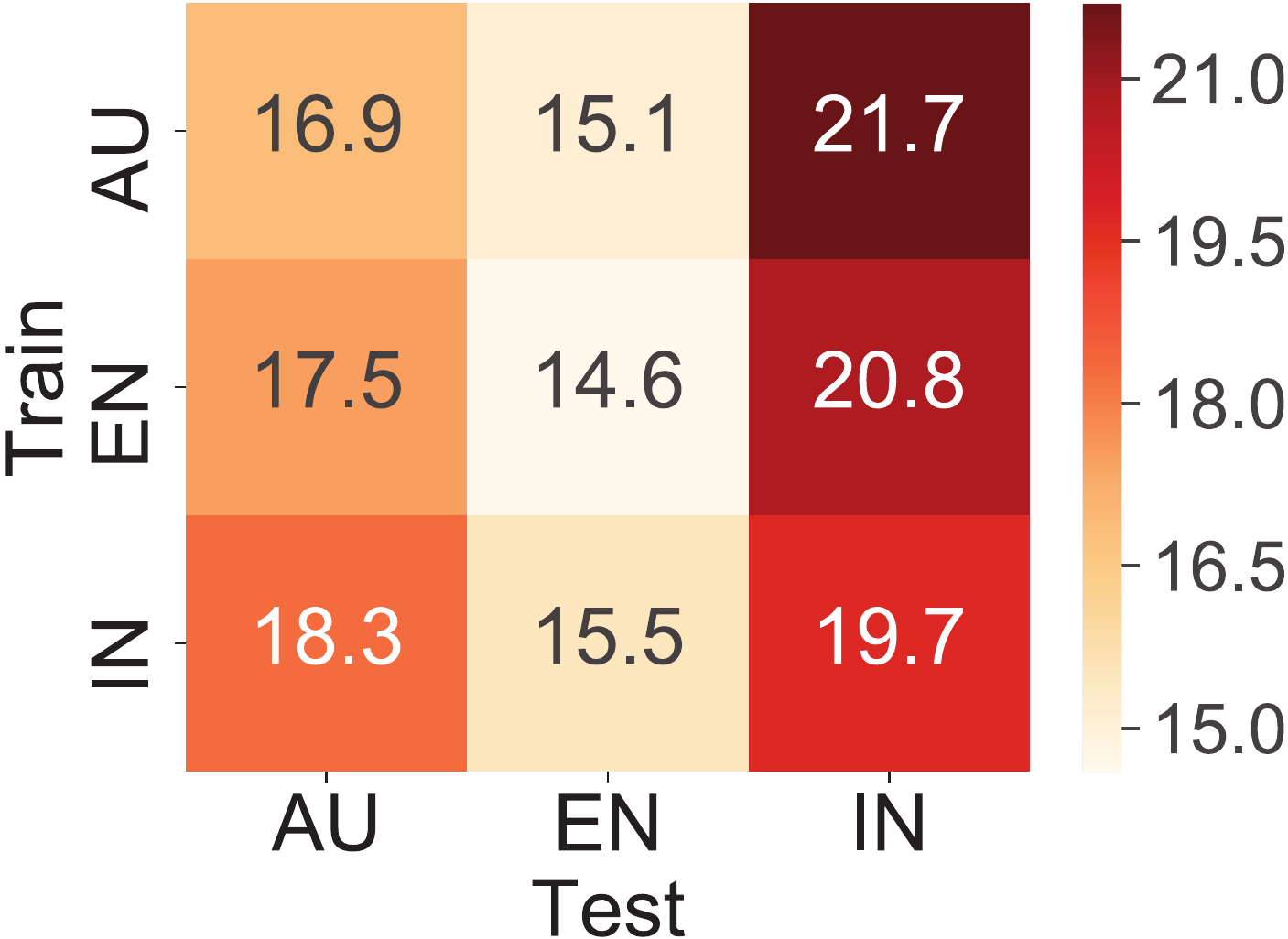}
	}
    \thinspace
	\subfloat[FedExtract]  
	{  \includegraphics[width=0.29\columnwidth]{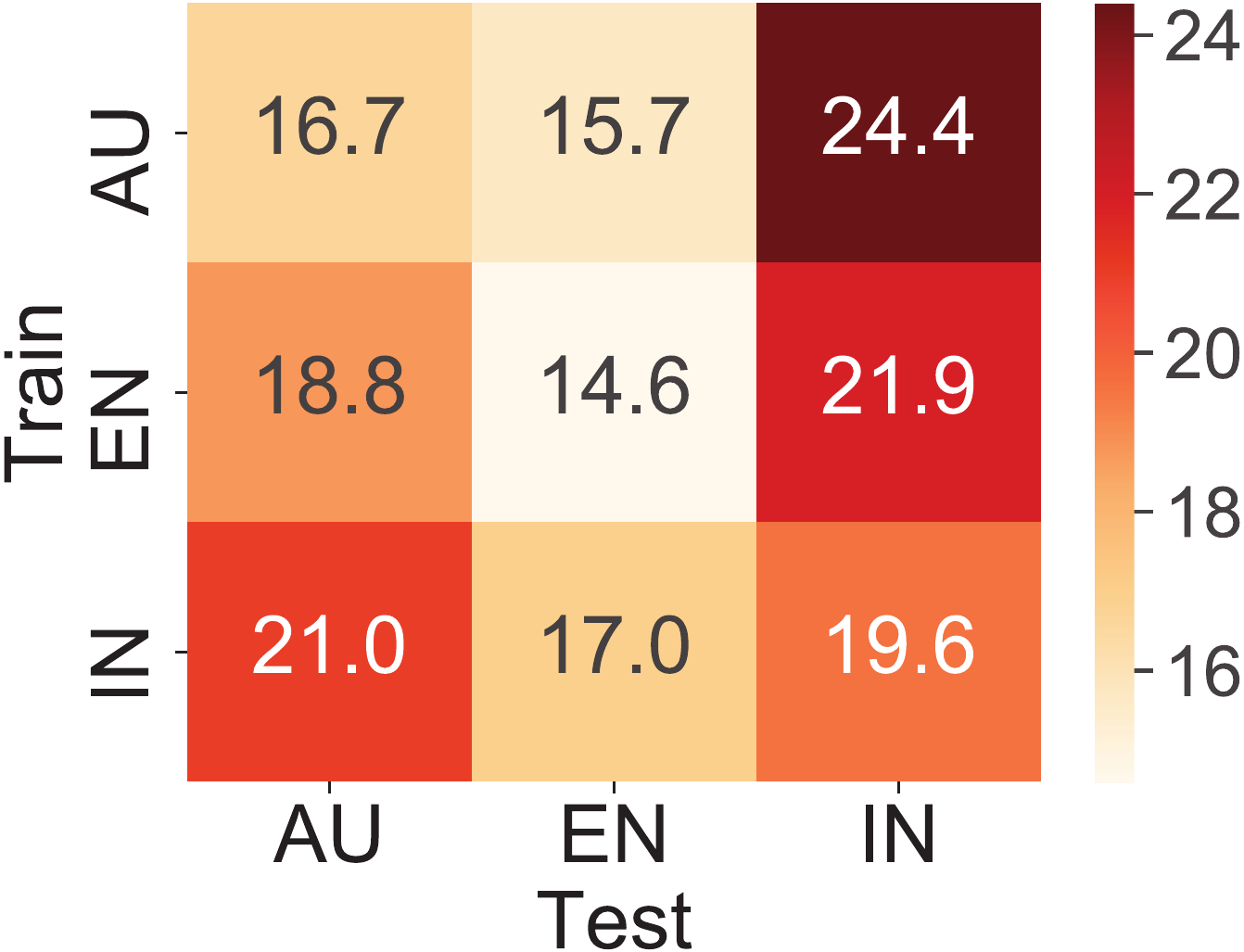}
	}
    \thinspace
	\subfloat[DecoupleFL]  
	{  \includegraphics[width=0.31\columnwidth]{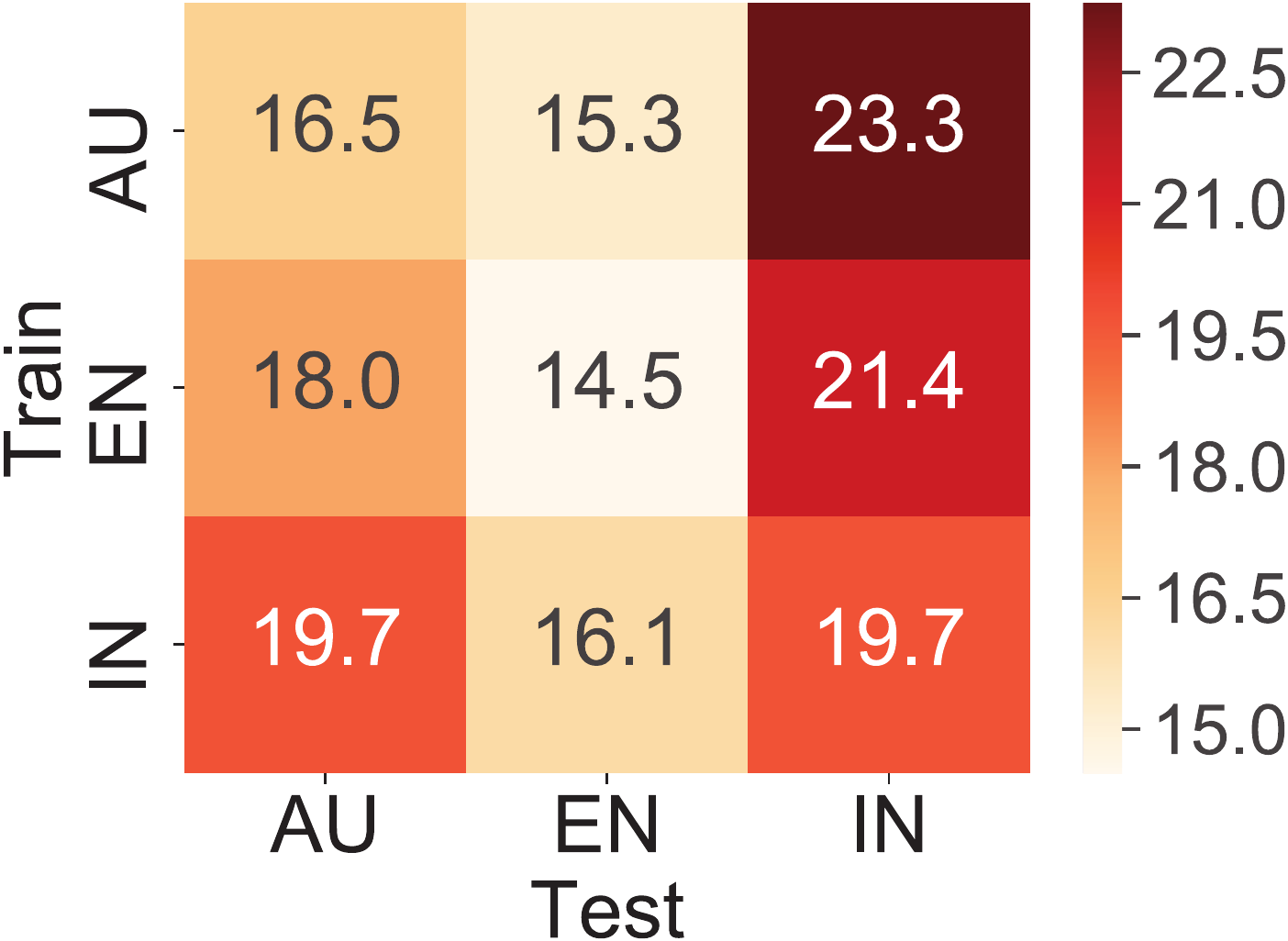}
	}
	\caption{Illustration of personalization effect.} 
	\label{fig:WER}
\end{figure}

We illustrate the personalization effects of three personalized FL approaches in \autoref{fig:WER}. The y-axis and x-axis denote the trained personalized client models and the client-specific testing set, respectively. All approaches perform best when we use the client model on this own testing set and degrades otherwise.

\subsection{Ablation Study of \method}
In this section, we perform the ablation study to show the importance of each component in \method. Note that each approach in the ablation study is trained for the same epochs.

\begin{table}[htbp]
  \centering
  \caption{Ablation study of \method}
  \resizebox{.25\textwidth}{!}{
    \begin{tabular}{lcccc}
    \toprule
    Method & AU & EN & IN & AVG \\
    \midrule
    Baseline & 19.4 & 15.7 & 22.6 & 19.2 \\
    \midrule
    \method & 16.5 & 14.5 & 19.7 & 16.9 \\
    - stage 1 & 17.3 & 14.7 & 20.3 & 17.4 \\
    - stage 2 & 16.7 & 14.9 & 20.3 & 17.3 \\
    - decouple & 16.8 & 14.5 & 19.6 & 17.0 \\
    \bottomrule
    \end{tabular}%
    }
  \label{tab:ablation}%
\end{table}%

As shown in \autoref{tab:ablation}, when stage 1 is discarded, the classifier is only trained with features extracted by a shared extractor. The performance degrades due to the lack of personalization.
On the other hand, when stage 2 is removed, each client optimizes a client-specific extractor while keeping the classifier fixed. The performance is also significantly degraded.
Finally, we discard decoupled training and perform end-to-end optimization for the entire model, which is equivalent to centralized training with personalized extractors for each client. The results show that the decoupled optimization can achieve similar performance to the end-to-end optimization.

\subsection{Privacy Protection in \method}
\label{sec:privacy}

\begin{figure}[!ht]
\centering
	\subfloat[CNN extractor] 
	{ \label{fig:tsne_conv}
		\fbox{\includegraphics[width=0.4\columnwidth]{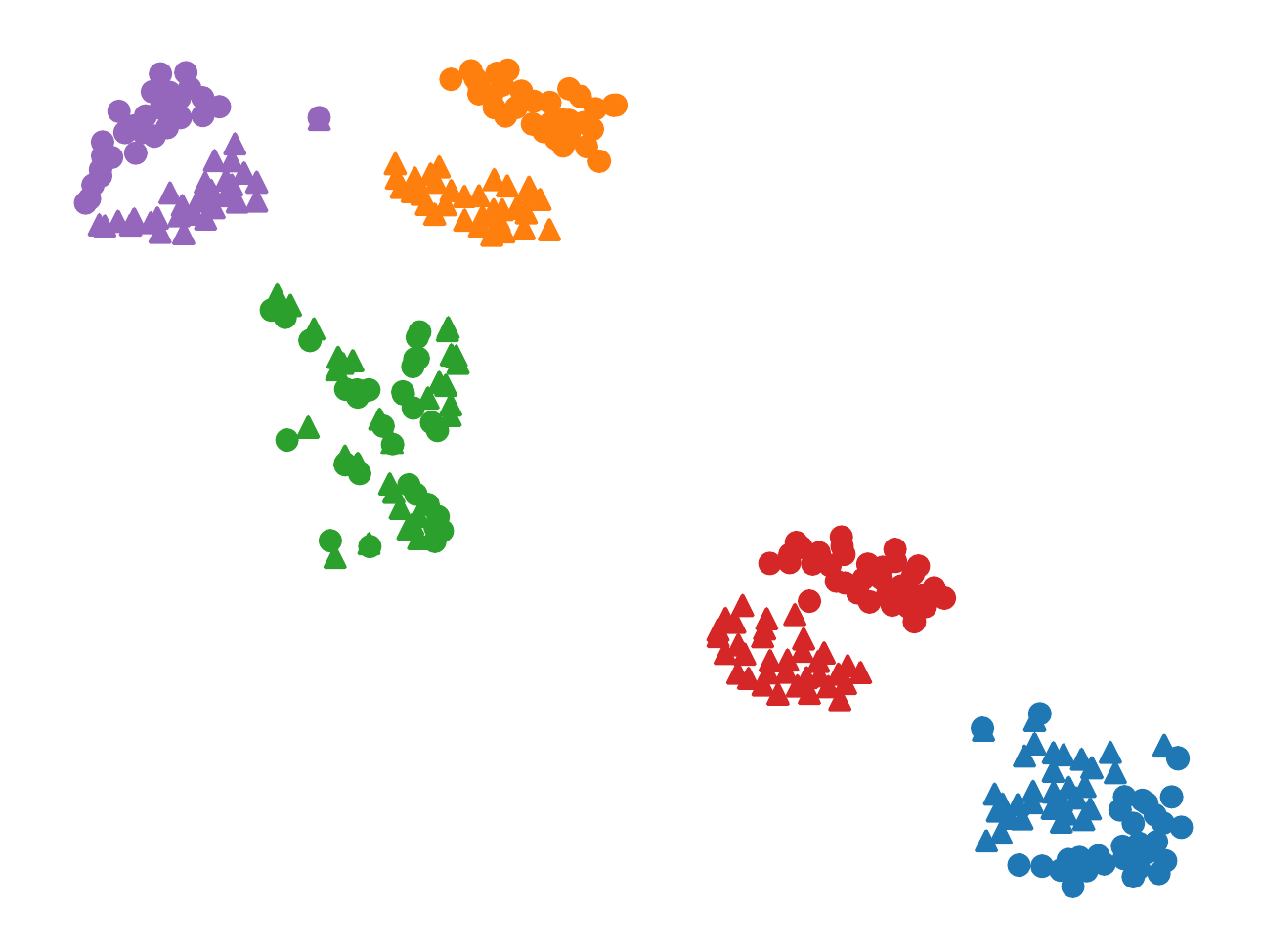}}
	}
	\thinspace
	\subfloat[\centering{CNN extractor} (SIT)] 
	{ \label{fig:tsne_conv_adv}
		\fbox{\includegraphics[width=0.4\columnwidth]{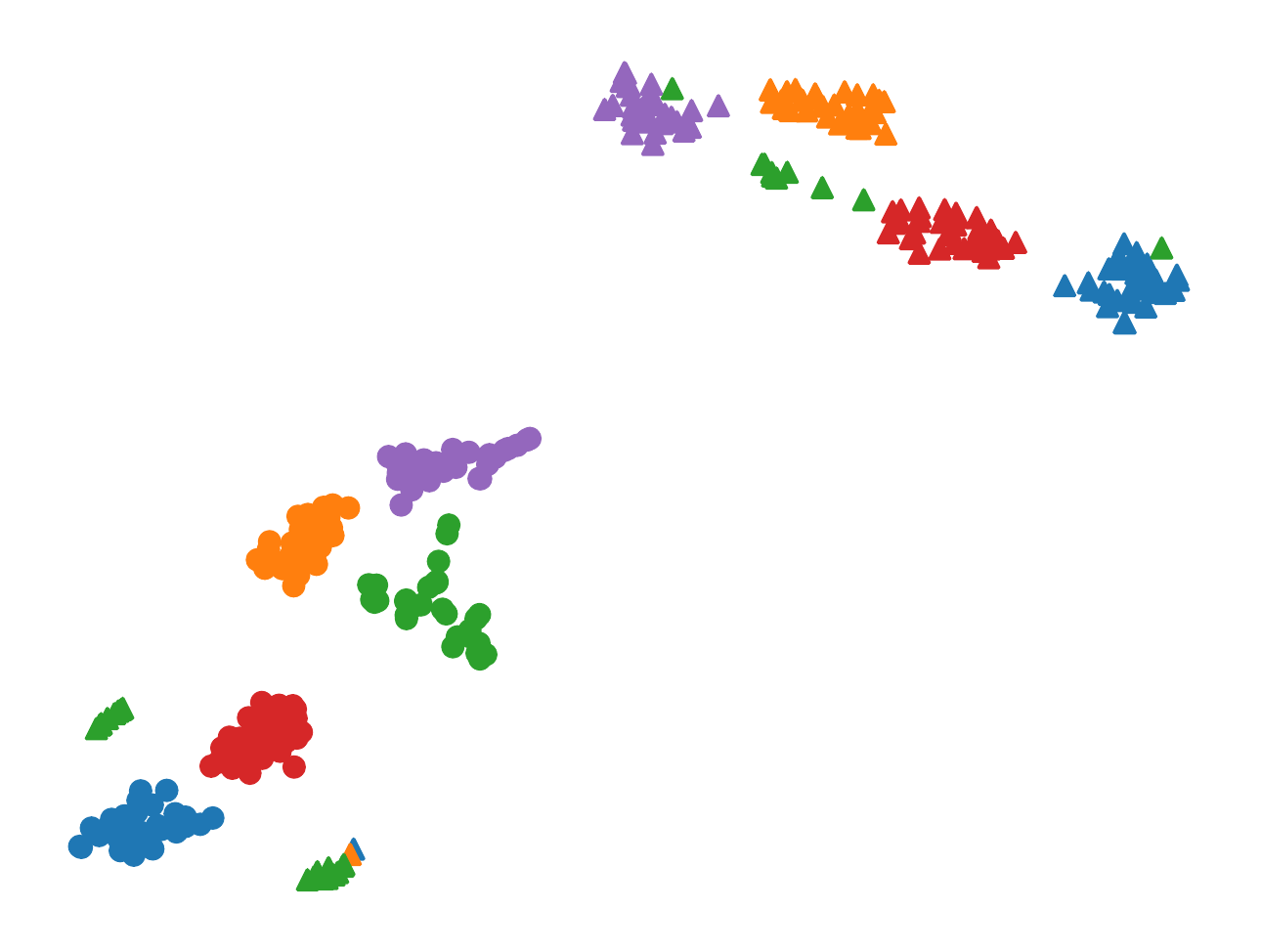}}
	}

	\subfloat[CNN+Transformer extractor] 
	{ \label{fig:dtsne_half}
		\fbox{\includegraphics[width=0.4\columnwidth]{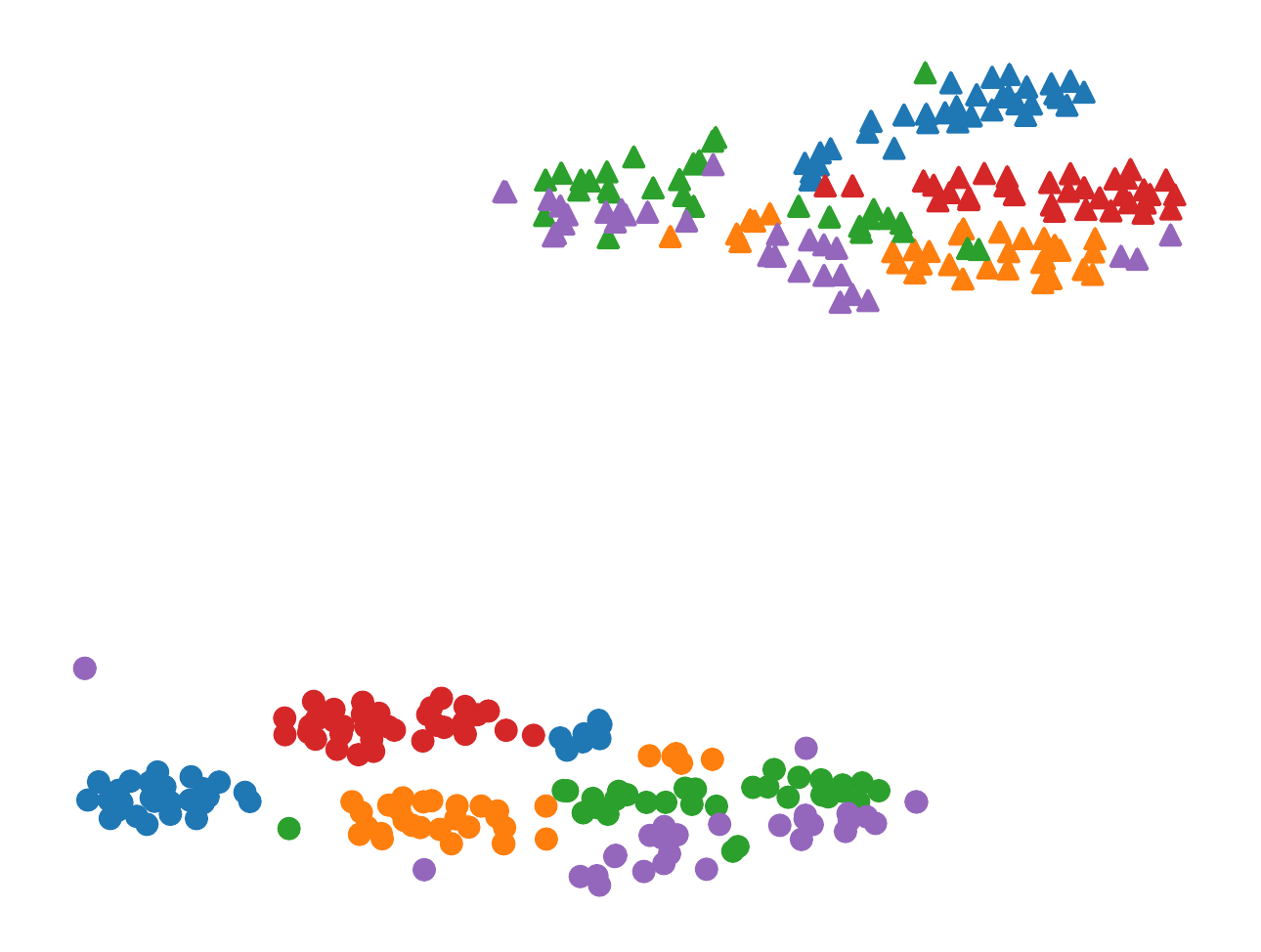}}
	}
	\thinspace
	\subfloat[\centering{CNN+Transformer extractor  (SIT)}] 
	{ \label{fig:tsne_half_adv}
		\fbox{\includegraphics[width=0.4\columnwidth]{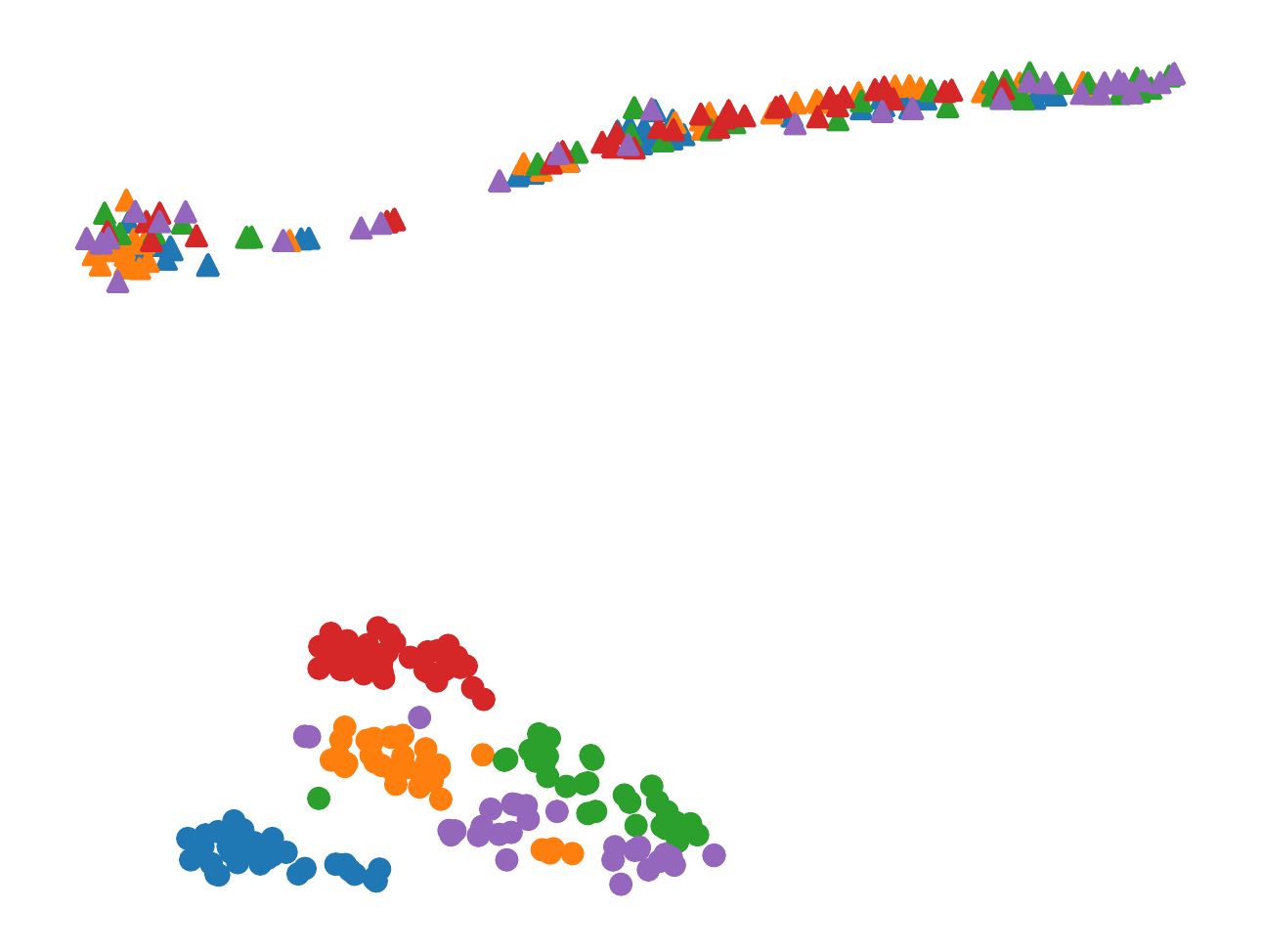}}
	}
	\caption{t-SNE Visualization of 5 speakers features from a client (EN). Unadapted and personalized features are denoted by circles and triangles, respectively. Each color represents a speaker. \emph{Best viewed in color and zoom in.}} 
	\label{fig:tsne}
\end{figure}

We illustrate the privacy protection properties for \method by discussing attacks against it.
The inversion attack \cite{fredrikson2015model} that aims to reconstruct the raw data could be avoided since the personalized extractors are only stored locally.
However, it is unclear whether the attribute inference attack \cite{feng2021attribute,tomashenko2021privacy} that tries to infer private attributes (like speaker identity) could be defended. We denote features extracted from unadapted and personalized extractors as \emph{unadapted} and \emph{personalized} features, respectively.
There are two potential attacking strategies to infer speaker information from personalized features: (\uppercase\expandafter{\romannumeral1}) an attacker who has audios of a speaker tries to retrieve this speaker's personalized features. Since the personalized extractors are invisible to the attacker, the attacker could only utilize the unadapted feature of this raw audio to find similar personalized features. To avoid this attack, the unadapted and personalized features of the same speaker should be sufficiently different. (\uppercase\expandafter{\romannumeral2}) the attacker tries to cluster the personalized features by speakers. To avoid this attack, the personalized features of the same speaker should not in the same neighborhood.

We illustrate the risks by visualizing unadapted and personalized features. We consider two types of extractors: CNN or CNN + bottom 7 transformer layers (CNN+Transformer). 

As shown in \autoref{fig:tsne}, with CNN extractor, unadapted and personalized features of the same speaker are close, indicating attack \uppercase\expandafter{\romannumeral1} is easy. SIT can help separate the unadapted and personalized features. However, the CNN extractor is not powerful enough to learn speaker-invariant representation. Thus the attack \uppercase\expandafter{\romannumeral2} is still possible.
With CNN+transformer extractor, the unadapted and personalized features are already separated without SIT, avoiding the attack \uppercase\expandafter{\romannumeral1}. And the personalized features become speaker-invariant with SIT, which avoids the attack \uppercase\expandafter{\romannumeral2}. Note that each client owns multiple speakers. Otherwise, the attack \uppercase\expandafter{\romannumeral2} could be done simply by tracing who sends the feature.

In conclusion, \method can protect data privacy since (1) Raw data does not leave clients; (2) Personalized extractors are stored in clients, thus avoiding the inversion attack; (3) SIT is used to protect the speaker identity in the features; (4) We could further improve the security of the features with homomorphic encryption \cite{rivest1978data} and differential privacy \cite{abadi2016deep}.

\subsection{Selection of Extractor}

We have illustrated that the CNN+Transformer extractor is better than CNN regarding privacy protection. Now we discuss the performance and computation of different extractors.

\begin{table}[htbp]
  \centering
  \caption{The performance of \method w. and w./o SIT exploiting different extractors}
    \resizebox{0.3\textwidth}{!}{
    \begin{tabular}{cccccc}
    \toprule
    Extractor & \multicolumn{1}{l}{SIT} & AU  & EN  & IN  & AVG \\
    \midrule
    \multirow{2}[2]{*}{CNN} & ×   & 16.7 & 14.6 & 19.7 & 17 \\
        & \Checkmark   & 16.7 & 14.8 & 19.7 & 17.1 \\
    \midrule
    \multirow{2}[2]{*}{CNN+Transformer} & ×   & 16.5 & 14.5 & 19.7 & 16.9 \\
        & \Checkmark   & 16.4 & 14.7 & 19.8 & 17.0 \\
    \bottomrule
    \end{tabular}%
    }
  \label{tab:spk_adv}%
\end{table}%

For performance, as shown in \autoref{tab:spk_adv}, SIT does not hamper the performance clearly. And the performance with CNN+Transformer extractor slightly outperforms the one with CNN extractor. 
In terms of computation, no matter which extractor we select, the computation is the same in stage 1 since the trainable extractor is in the bottom of the model, which requires the forward and backpropagation for the entire model. Although the training memory in stage 1 could be decreased with a smaller extractor since only gradients of the extractor are kept in memory, the computation will correspondingly increase in stage 2 since a larger classifier is trained. 
Therefore, it is reasonable to use CNN+Transformer rather than CNN as the extractor for better privacy, performance and computation.

\section{Discussions}
\label{sec:discussion}

We discuss the proposed approaches as follows.

\textbf{Data privacy.} Personalization layer based FL, i.e., FedNorm and FedExtract, follow the FedAvg paradigm, whose security property is thoroughly discussed in previous work~\cite{tomashenko2021privacy,fowl2021robbing}. For \method, we illustrate that it could protect privacy in \autoref{sec:privacy} and more discussions are required in future. 
    
\textbf{Communication cost.} \method transfers features once, while FedNorm and FedExtract transfer model parameters multiple times. Therefore, when data size per client is relatively small or the global epochs are large, \method will have less communication cost and vice versa.

\textbf{Computation cost.} FedNorm and FedExtract perform training on clients and the computation on the server could be neglected. But \method trains on clients for half epochs (stage 1) and transfers the other half to the server (stage 2). In the server, since we only train the classifier, computation is reduced by 50\%, which results in the 25\% reduction in total.

\section{Conclusions}
\label{sec:conclusion}

In this work, we tackle FL-based ASR in the non-IID scenario with personalized FL. Firstly, We adapt the personalization layer based personalized FL approach for ASR. Secondly, we propose \method to reduce communication and computation costs. Experiments show that the two types of personalized FL approaches achieve lower WER than FedAvg. Among them, \method has the lowest communication and computation cost. Furthermore, we show \method can protect data privacy with SIT without performance compromise.

\bibliographystyle{IEEEtran}
\bibliography{./main}

\end{document}